\documentclass{KapProc} 

\input epsf

\let\footnote\savefootnote

\let\footnotetext\savefootnotetext 

\let\footnoterule\savefootnoterule

\kluwerbib

\begin{document}

\articletitle[Massive stellar winds]
{Stellar Winds from Massive Stars}

\author{Paul A. Crowther}

\affil{Department of Physics \& Astronomy, University College London,
Gower St., London WC1E 6BT, U.K.} 

\email{pac@star.ucl.ac.uk}


\begin{abstract}
We review the various techniques through which 
wind properties of massive stars --
O stars, AB supergiants, Luminous Blue Variables (LBVs), 
Wolf-Rayet (WR) stars and cool supergiants -- are derived.
The wind momentum-luminosity relation (e.g. Kudritzki et
al. 1999) provides a method
of predicting mass-loss rates of O stars and blue supergiants
which is superior to previous parameterizations.
Assuming the theoretical $Z^{0.5}$ metallicity dependence,  
Magellanic Cloud O star mass-loss rates are typically matched to
within a factor of two for various calibrations. 
Stellar winds from LBVs are typically denser 
and slower than equivalent B supergiants, with 
exceptional mass-loss rates during giant eruptions
$\dot{M}=10^{-3} \ldots 10^{-1} M_{\odot}$ yr$^{-1}$ (Drissen et al. 2001). 
Recent mass-loss rates for Galactic WR stars 
 indicate a downward revision of 2--4 
relative to previous calibrations due to clumping (e.g. Schmutz 1997), 
although evidence for
a metallicity dependence remains inconclusive (Crowther 2000). 
Mass-loss properties of luminous ($\ge 10^{5} L_{\odot}$) 
yellow and red supergiants from alternative techniques remain
highly contradictory. Recent Galactic and LMC results for RSG 
reveal a large scatter such that typical mass-loss rates lie in the 
range $10^{-6} \ldots 10^{-4} M_{\odot}$ yr$^{-1}$, with a 
few cases exhibiting $\sim 10^{-3} M_{\odot}$ yr$^{-1}$.
\end{abstract}

\section{Introduction}

Winds are ubiquitous in massive stars, although the physical 
processes involved  depend upon the
location of the star within the H-R diagram. Mass-loss crucially
affects the evolution and fate of a massive star (Meynet
et al. 1994), while the 
momentum and energy expelled contribute to the dynamics and 
energetics of the ISM. The interested reader is referred to Lamers \& 
Cassinelli (1999) for the topic of stellar winds in general, or 
Kudritzki \& Puls (2000) for a detailed discussion of mass-loss from OB
 stars. This review will cover the broad topics of mass-loss from
early-type (OBA, Wolf-Rayet, Luminous Blue Variable) and late-type 
(yellow/red supergiants) stars.  
Theoretical predictions for hot star winds follow line-driven
radiation pressure (Castor et al. 1975; Pauldrach et al. 1986). 
Pulsations are thought to initiate the winds of cool supergiants, together
with other mechanisms (e.g. Alfven waves). Radiation pressure on dust grains
helps to maintain outflows in their cool, outer envelopes (Wickramasinghe
et al. 1966).

\section{Hot Star wind diagnostics}\label{ob}

Global wind properties can be characterized by mass-loss and wind velocity.
In hot stars, the former depends on application of varying complexity of
theoretical interpretation, while the latter can fortunately be 
directly measured with minimal interpretation. 

\begin{figure}[ht]
\epsfxsize=11cm \epsfbox[0 320 550 780]{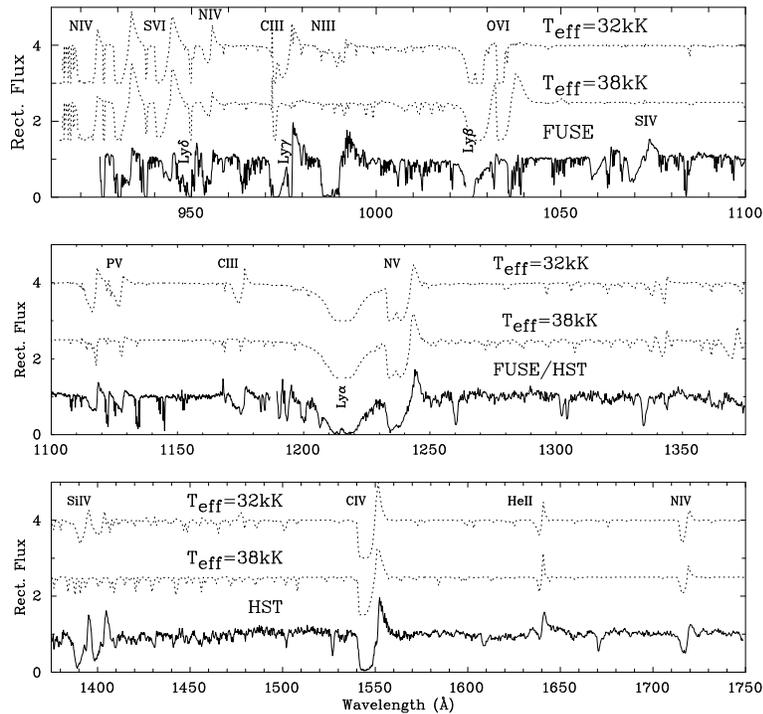}
\caption{Comparison between HST UV and FUSE far-UV spectra of 
Sk\,80 (O7\,Iaf) with synthetic spectra (including shocks) 
at  $T_{\rm eff}$=37,500K
and 32,000K, illustrating the improved agreement for the latter value
(Fullerton et al. 2000).}
\label{FIG1}
\end{figure}

\subsection{UV and far-UV spectroscopy}\label{uv}

Ultraviolet P Cygni profiles, ubiquitous in O-type stars (Walborn
et al. 1985) provide a direct indication of stellar winds.
Such metal resonance lines are generally analysed via the Sobolev 
with exact integration (SEI) method  (Lamers et al. 1987; Haser
et al. 1998). 
Mass-loss determinations require knowledge of
elemental abundances, the degree of ionization of the
ion producing the line, and an assumed form of the velocity law 
that is predicted by radiation driven winds. 
In practice, accurate wind velocities can
be readily obtained from saturated UV P Cygni profiles 
(Groenwegen et al. 1989; Prinja et al. 1990), while unsaturated 
P Cygni profiles provide a better means
of determining mass-loss rates, albeit still subject to ionization 
results from non-LTE atmospheric models. Such models are
subject to uncertain contributions of ionizing radiation from 
shocks in the stellar outflow. In view of these difficulties, 
Lamers et al. (1999) have combined column densities from unsaturated UV
P Cygni profiles with independently determined mass-loss rates (from
radio or H$\alpha$, see below) 
to derive empirical ionization fractions. 

The Far-Ultraviolet Spectroscopic Explorer (FUSE) offers the possibility 
of determining mass-loss rates solely from UV data alone, since the far-UV
significantly increases the number of stellar wind diagnostics, including
C\,{\sc iii}, N\,{\sc iii-iv}, O\,{\sc vi}, S\,{\sc iv-vi} and P\,{\sc v}.
The additional ionization information 
provided by FUSE permits stellar temperature determinations. Fullerton
et al. derived $T_{\rm eff}$=32--33kK for two Magellanic Cloud O7 
supergiants from UV wind metal diagnostics, 
in contrast to $T_{\rm eff}$=38kK from the usual optical helium
plane-parallel hydrostatic methods (see Fig.~\ref{FIG1}). 
Although previously derived
mass-loss rates were confirmed, lower bolometric corrections implied
reduced stellar luminosities, with consequences for current 
mass-loss calibrations.  

\subsection{Optical and near-IR spectroscopy}

Although H$\alpha$ has long been recognized as the prime
source of mass-loss information in early-type stars,
accurate determinations rely upon a complex treatment of the lines and
continua, i.e. non-LTE, spherically extended 
models which treat the sub- and supersonic atmospheric structure.
Use of such codes also provide surface temperatures and gravity in 
OB stars via optical photospheric lines (e.g. He\,{\sc i-ii} in O stars).
Problems with using fits to H$\alpha$ to derive mass-loss
rates include blending with He\,{\sc ii} $\lambda$6560 
in O stars, together with
nebular contamination from H\,{\sc ii} regions (e.g. NGC346; 
Walborn et al. 1995) and broadening of the central emission
component by stellar rotation.
In A supergiants, H$\alpha$ behaves as a scattering line, with
a characteristic P Cygni profile, so they have the advantage over 
other hot stars that 
wind velocities and mass-loss rates may be simultaneously
determined in external galaxies without the need for UV spectroscopy.

Although moderate to high spectral resolution observations of OB stars
are rare at near-IR wavelengths (e.g. Fullerton \& Najarro 1998),
analogous lines to optical wind (and photospheric) diagnostics are 
available (Bohannan \& Crowther 1999). This provides the prospect of
the determination of mass-loss properties for hot stars obscured at 
optical wavelengths via IR diagnostics (Najarro et al. 1994).

\subsection{IR--radio continua}

Winds in hot stars can be readily observed at IR-mm-radio wavelengths
via the free-free (Bremsstrahlung) `thermal' excess caused by the 
stellar wind, under the assumption of homogeneity and spherical symmetry.
Mass-loss rates can be determined via application of relatively simple
analytical relations (Wright \& Barlow 1975; Panagia \& Felli 1975).
Barlow \& Cohen (1977) have used IR excess fluxes to determine 
mass-loss rates for a large sample of 
Galactic OBA supergiants, which generally compare
favourably 
with recent H$\alpha$ results for A and B supergiants (Kudritzki et al. 
1999). Unfortunately, OB stars with relatively weak winds do not show
a strong IR excess or radio flux,  so mass-loss results have solely been for 
nearby hot stars with dense winds (Bieging et al. 1989; Drake \& Linsky 1989).
Observations at multiple frequencies are necessary to ensure against 
non-thermal (synchrotron) radio emission from colliding winds (e.g. Chapman 
et al. 1999). H$\alpha$ and radio continuum mass-loss rates agree
reasonably well (Lamers \& Leitherer 1993) despite sampling quite different
parts of the stellar wind -- $\sim$1.5 stellar radii for H$\alpha$ and
$\sim$1000 stellar radii for the radio photosphere -- placing constraints
on relative clumping factors in these regions.

\section{Results for OBA stars - role of metallicity?}

Although the techniques used to study hot, massive stars are similar,
we first discuss `normal' early-type stars, and leave `exotica' until 
subsequent sections. 
The mass-loss rate of a star is a fundamental ingredient in evolutionary
models. Unfortunately,  radiatively
driven wind theory in its current form (Puls et al. 1996; Vink
et al. 2000) is unable to
reproduce the observed mass-loss properties for hot luminous
stars sufficiently accurately (see below). Nevertheless, since
their winds are driven by photon momentum transfer through metal line 
absorption, the stellar momenta and wind velocities are expected
to be functions of stellar metallicity, $Z$, such that mass-loss rates
scale with $Z^{0.5}$ for O stars and early B stars,
$Z^{0.8}$ for mid-B supergiants, and $Z^{1.7}$ for 
A supergiants (Kudritzki \& Puls 2000). 

\subsection{Terminal Wind Velocities}

Large compilations of wind velocities, $v_{\infty}$, 
of O-stars and AB supergiants 
have been made by Prinja et al. (1990) and Lamers et al. (1995), revealing
a spectral type and luminosity class dependence. Lamers et al. highlighted
the so-called `bi-stability' jump in wind properties around B\,1.5 
($\sim$21,000), 
above which  $v_{\infty} \simeq 2.65 v_{\rm esc}$ (escape velocity), 
and $v_{\infty} \simeq 1.4 
v_{\rm esc}$ below, resulting from the change in ionization of the
elements contributing to the line force (Vink et al. 1999).
Wind velocities of LMC
O stars differ little from Galactic counterparts (Garmany \& Conti 1985), 
while a more prominent effect is observed in the SMC, 
particularly amongst early O stars (Walborn et al. 1995;
Prinja \& Crowther 1998).

\begin{table}[ht]
\caption{Coefficients for wind momentum luminosity relationship 
of Galactic O-stars and AB supergiants (Kudritzki \& Puls 2000)}
\label{wmr}
\begin{center}
\begin{tabular}{lcc}
\hline
\it Sp Type & \it log $D_{0}$ & {\it $x$} \cr
\hline
O dwarf/giant    & 19.87$\pm$1.21 & 1.57$\pm$0.21 \cr
O supergiant     & 20.69$\pm$1.04 & 1.51$\pm$0.18 \cr
B0--1 supergiant & 21.24$\pm$1.38 & 1.34$\pm$0.25 \cr
B1.5--3 supergiant  & 17.07$\pm$1.05 & 1.95$\pm$0.20 \cr
A0--3 supergiant     & 14.22$\pm$2.41 & 2.64$\pm$0.47 \cr
\hline
\end{tabular}
\end{center}
\end{table}

\subsection{Mass-loss rates}

 Observationally, estimates of mass-loss rates 
from H$\alpha$, UV wind lines  or radio fluxes have been available for 
the past 20 years, although  is only recently that a reasonable 
degree of consistency  has been reached 
between these different approaches via theoretical improvements.
Puls et al. (1996) and Kudritzki et al. (1999) identified that O-stars
and AB supergiants obey a wind momentum-luminosity relationship, as follows

\[ \log  \dot{M} v_{\infty} (R/R_{\odot})^{0.5} = \log D_{0} + x \log (L/L_{\odot}) \]

Stars of different spectral types have differing $D_{0}$ and $x$ 
values, listed in Table~\ref{wmr}, because
of the change of lines driving the stellar wind with $T_{\rm eff}$.

Hot stars in the Magellanic Clouds provide us with the means to compare
emirical mass-loss properties with predictions at lower $Z$
(e.g. Garmany \& Conti 1985). This has proved to be
difficult given (i) the inability to measure radio fluxes for extra-galactic 
hot stars; (ii) the need to combine high quality UV and optical spectroscopy 
with sophisticated model techniques; (iii) the limited metallicity range
spanned, 0.15--1$Z_{\odot}$. 
Although the number of stars analysed in detail for the Magellanic Clouds
remains small (Puls et al. 1996), the mass-loss rates of LMC and SMC 
O stars tend to be lower than Galactic counterparts. 
As an example, let us consider Sk\,80 (O7\,Iaf+) in the SMC giant
H\,{\sc ii} region NGC\,346 which has a metallicity of 0.2$Z_{\odot}$ 
(Haser et al. 1998; Peimbert et al. 2000).  From the wind 
momentum-luminosity relationship we would expect an equivalent 
Galactic supergiant to have a mass loss rate four times higher 
than its observed value (Puls et al. 1996), suggesting
a higher exponent of $Z^{0.8}$ in this case.

\begin{table}[ht]
\caption{Comparison ($\dot{M}(T_{\rm eff}, L)/\dot{M}_{\rm obs}$; 
$\sigma$ in parenthesis)
between predicted OBA mass-loss rates from
various parameterizations with empirical data from Puls et al. (1996)
and Kudritzki et al. (1999). Parameterizations are
de Jager et al. 1988; Lamers \& Cassinelli 1996;
Vanbeveren et al. 1998; Kudritzki \& Puls (2000); Vink et al. (2000) or
Achmad et al. (1997) for A supergiants. Except for
Lamers \& Cassinelli, no metallicity dependence is accounted for,
so we here assume a metallicity dependence of $Z^{0.5}$ for O stars,
with 0.4$Z_{\odot}$ for the LMC, 0.15$Z_{\odot}$ for the SMC.}
\label{mdot}
\begin{center}
\begin{tabular}{l@{\hspace{2mm}}l@{\hspace{2mm}}l@{\hspace{2mm}}l
@{\hspace{2mm}}l@{\hspace{2mm}}l@{\hspace{2mm}}l}
\hline
\it Spect.      & \it Galaxy & \it de Jager & \it Lamers \&&
\it Vanbeveren & \it Kudritzki & \it Vink  \cr
\it type (\#)  &        & \it et al.    &
\it Cassinelli&\it et al.  & \& \it Puls & \it et al. \cr
\hline
O (23)&Gal&0.9 (0.7)&\phantom{1}1.1 (0.6)&0.9 (0.7)&1.3 (0.8)  &\phantom{1}1.8 
\phantom{1}(1.3)\cr
O \phantom{1}(6) &LMC&0.5 (0.5)&\phantom{1}1.3 (0.7)&0.5 (0.3)&0.7 (0.3)  &\phantom{1}1.3 
\phantom{1}(0.4)\cr
O \phantom{1}(5) &SMC&1.3 (0.9)&\phantom{1}2.5 (1.5)&1.3 (0.9)&2.3 (1.3)  &\phantom{1}6.5 
\phantom{1}(7.9)\cr
B (14)&Gal&6.8 (4.6) &11.3 (7.7)&        &1.0 (0.3)&38.8 (29.4)\cr
A \phantom{1}(4) &Gal&5.0 (1.6) &\phantom{1}3.7 (1.9)&        &1.0 (0.3)  &\phantom{1}2.0 \phantom{1}(1.0)\cr
\hline
\end{tabular}
\end{center}
\end{table}

Evolutionary calculations require  $\dot{M}(T_{\rm eff}, L)$ parameterizations 
based on empirical data. Up until recently, the only compilation
covering hot, luminous stars was that of de Jager et al. (1988). 
In an attempt to update the mass-loss calibration for evolutionary 
calculations,  Vanbeveren et al. (1998) provided an 
updated relationship for Galactic OB stars, while 
Lamers \& Cassinelli (1996) additionally considered LMC/SMC O stars. 
Most recently, Achmad et al. (1997) and Vink et al. (2000) have provided
a set of predictions from current radiative driven winds for 
A, F and G supergiants and OB stars, respectively. How do these different  
parameterizations compare with observed mass-loss rates?

We have taken empirical data, $\dot{M}_{\rm obs}$, 
from Puls et al. (1996) for O stars and Kudritzki et al. (1999) for 
AB supergiants. In Table~\ref{mdot} we compare the mean ratio 
$\dot{M}(T_{\rm eff}, L)/\dot{M}_{\rm obs}$ for each calibration.
Perhaps surprisingly, de Jager et al. (1988) fares as well as the recent 
Lamers \& Cassinelli or Vanbeveren et al. (1998) parameterizations for O 
stars.  In contrast with previous  theoretical results, 
Vink et al. (2000)  predicts sufficiently strong winds, albeit
a factor of 2 too {\it strong} for Galactic O stars. 
Overall, the mass-loss rates of AB supergiants are typically
overestimated (by 5--10) with the exception of 
Kudritzki et al. (1999) and Achmad et al. (1997). Predictions by
Vink et al. (2000) show the poorest agreement for mid-B supergiants.

Overall, solely the wind momentum-luminosity prescription 
provides excellent agreement with all OBA observations. One note of 
caution is warranted, since the $T_{\rm eff}$-calibration for 
O stars, and hence the luminosity scale, relies almost universally 
on results from plane-parallel studies which neglect stellar winds
(recall Sect~\ref{uv}).

\subsection{Rotation}

O star rotational velocities ($v \sin i$) are typically 100 km\,s$^{-1}$
(Howarth et al. 1997), with important consequences for their 
evolution (Meynet \& Maeder 2000). 
Friend \& Abbott (1986) first considered the effect of
centrifugal acceleration on hot star winds, extended by Bjorkman \& 
Cassinelli (1993) to allow for the azimuthal dependence, who found
that $\dot{M}$ increases (and $v_{\infty}$ decreases) towards the equator,
leading to an oblate `wind-compressed disk'. 

Subsequently, Owocki et al. (1996) established that quite different
predictions may result, by additionally allowing for non-radial 
components of the line force, and the increased polar radiation flux 
via gravity darkening (von Zeipel 1924). Due to a reduced escape velocity, the 
equatorial velocity law is slower than at the pole, inhibiting the 
equatorial disk and the original scaling of mass-loss can be reversed, such 
that a prolate geometry results. As an example, a mid-B dwarf rotating 
at 85\% of its critical velocity was found by Petrenz \& Puls (2000) 
to have a strongly prolate structure ($\rho_{\rm pole}/\rho_{\rm eq}
\leq$15) with a polar/equatorial terminal velocity of 1030/730
km\,s$^{-1}$. The global mass-loss rate was not found to deviate from
its 1D value by  greater than 10--20\%, except for supergiants close 
to the Eddington limit.

\subsection{Structure}

Observationally, evidence for structure in O stars winds has
taken various forms, including absorption variability observed from
optical (Fullerton et al. 1996) and UV (Prinja 1992) spectroscopic 
intensive monitoring, and the presence of soft X-ray  emission (Chlebowski
et al. 1989), attributed to shocks in their winds (Lucy \& White 1980). 
This view has recently been questioned by Chandra observations of
$\zeta$ Ori (Waldron \& Cassinelli 2000) which reveal that the X-ray 
emitting gas is exceptionally dense, with no evidence for expansion.

Theoretical studies of 
hot star winds reveal that line driven acceleration is subject to a 
strong instability, causing a variable and structured winds (Owocki et al.
1988; Feldmeier 1995). Since the region where H$\alpha$ forms in OB stars 
($\leq 1.5 R_{\ast}$) is not anticipated to be heavily structured, mass-loss 
determinations may not be greatly affected by clumping/structure. 
Recent extensive monitoring of HD\,152408 (WN9ha or O8\,Iafpe) 
by Prinja et al. (2000) indicate variations of $\pm$10\% in global 
mass-loss rate, although with a time averaged H$\alpha$ profile 
that is remarkably stable over several years.

\section{Luminous Blue Variables}

The stellar wind properties of LBVs are poorly constrained, although
it is apparent that their winds are generally slower and denser than
 OBA supergiants (Crowther 1997), with little variation in mass-loss
during their `normal' irregular excursions 
across the H-R diagram (Leitherer 1997). To illustrate this, let us 
consider P Cygni, whose parameters have been determined by Najarro et al. 
(1997). Compared to normal B1.5 supergiants (780 km\,s$^{-1}$; 
Lamers et al. 1995), P Cyg has a very slow wind (185 km\,s$^{-1}$) and
a mass-loss rate which exceeds the Galactic wind momentum relation
(Sect.~\ref{ob}) by a factor of 15. Recall that from Petrenz \& Puls (2000), 
non-rotating models of  supergiants close to the Eddington limit may 
overestimate global mass-loss rates by a factor of $\sim$2, so that 
differences may be somewhat less severe. There are numerous examples of
extreme supergiants which are not LBVs yet show remarkably 
high mass-loss rates and low wind velocities (e.g. Hillier et al. 1998).

In the case of giant LBV eruptions, information is very scarce and up until
recently solely reliant on the 1840's $\eta$ Car eruption causing the 
formation of the Homunculus, in which $\sim$3 $M_{\odot}$ was ejected 
over $\sim$30 years (Davidson \& Humphreys 1997), corresponding to 0.1 $M_{\odot}$
yr$^{-1}$. Within the past few years, two extra-galactic LBVs have had
sudden giant eruptions, HD\,5980 in the SMC (Barba et al. 1995) and 
V1 in NGC\,2363 (Drissen et al. 1997). These eruptions, in stars
at low metallicity environments, appear to be quite different from that
of $\eta$ Car during its eruption, since peak
mass-loss rates of each are $\sim 10^{-3} M_{\odot}$yr$^{-1}$ (Drissen 
et al. 2001), equivalent to the current mass-loss for $\eta$ Car (Hillier
et al. 2000).

\section{Wolf-Rayet stars}

The diagnostics of mass-loss for Wolf-Rayet stars mimic
O stars except that their stronger winds mean that 
UV/optical wind diagnostics are not restricted to resonance lines or 
H$\alpha$. 
Wind velocities can be accurately determined either from the usual
UV P~Cygni profiles (Prinja et al. 1990), or from optical/IR
He\,{\sc i} P Cygni profiles (Eenens \& Williams 1994).
Radio and mm fluxes have been used to determine mass-loss 
rates of Galactic WR stars (Leitherer et al. 1997), while  recent 
theoretical progress (Schmutz 1997; Hillier \& Miller 1998) 
now permits detailed non-LTE analysis of UV and optical observations.
WR winds are typically spherical (Harries et al. 1998), but clumped
(Lepine et al. 2000). Spectroscopic tools make use of electron
scattering wings in WR winds to estimate clumping factors since their
strength depends linearly 
on density, versus the square of density for the underlying
recombination line (Hillier 1991).
Typical volume filling factors are 5--10\%, such that mass-loss rates in
a clumpy medium are 3--4 times lower than those from 
homogeneous models (e.g. Hillier \& Miller 1999).

Although great progress has been made in recent years, detailed analysis
requires labour intensive use of complex models, so the sample of stars
studied in detail remains small. Recently, Nugis \& Lamers (2000) 
combined empirical IR-radio observations  with analytical models 
accounting for clumping to estimate mass-loss rates as a function of
luminosity, He ($Y$) and metal ($Z$) mass fractions, i.e.

\[ \log (\dot{M}_{\rm WN}) = -13.6 + 1.63 \log L/L_{\odot} + 2.22 \log Y \]

\[ \log (\dot{M}_{\rm WC}) = -8.3 + 0.84 \log L/L_{\odot} + 2.04 \log Y + 1.04 \log Z \]

Recent evolutionary
calculations for hydrogen-poor WR stars follow 
mass-loss calibrations of  Langer (1989) or Vanbeveren et al. (1998). 
Table~\ref{wr-cal} compares predictions from these calibrations with 
results from detailed non-LTE line blanketed spectroscopic analyses,
assuming the Schaerer \& Maeder (1992) mass-luminosity relationship.
The formulation of Langer (1989) overestimates mass-loss rates by factors
of 2--4, while Vanbeveren et al. (1998) and especially Nugis \& Lamers
(2000) show much improved agreement with spectroscopic results.

\begin{table}[ht]
\caption{Comparison between WR mass-loss rates, in 10$^{-5} 
M_{\odot}$yr$^{-1}$, predicted by Langer (1989, L89), Vanbeveren et al.
(1998, VDV98),  semi-empirical calibration of Nugis \& Lamers
(2000, NL2000) with detailed non-LTE spectroscopic results (nLTE) 
of hydrogen poor Galactic WN and WC stars.}
\label{wr-cal}
\begin{center}
\begin{tabular}{l@{\hspace{2mm}}l@{\hspace{2mm}}c@{\hspace{2mm}}c
@{\hspace{2mm}}c@{\hspace{2mm}}c@{\hspace{2mm}}c}
\hline
\it Star & \it Sp Type & \it L89 &\it VDV98 & 
\it NL2000 &\it nLTE & \it Ref\cr
\hline
WR6    & WN4    & 12.5 & 5.5 & 5.7 & 3.2 & a \cr 
WR147  & WN8(h) & \phantom{1}9.5 & 4.5 & 3.2 & 2.5 & b \cr 
WR111  & WC5    & \phantom{1}2.8 & 2.0 & 1.3 & 1.5 & c \cr 
WR90   & WC7    & \phantom{1}4.8 & 3.2 & 2.6 & 2.5 & d \cr 
WR135  & WC8    & \phantom{1}2.0 & 1.6 & 1.6 & 1.2 & d \cr 
\hline
\end{tabular}
\end{center}
\begin{tablenotes}
($a$) Schmutz (1997); ($b$) Morris et al. (2000); ($c$) 
Hillier \& Miller (1999); ($d$) Dessart et al. (2000)
\end{tablenotes}
\end{table}

The question of a $Z$--dependence of mass-loss in WR stars remains
open. There is a clear distinction amongst spectral subtypes in
galaxies of differing metal content, which has been qualitatively 
explained from evolutionary and spectroscopic
models (Smith \& Maeder 1991; Crowther 2000). 
LMC WN stars (Crowther \& Smith 1997; Hamann \& Koesterke 2000) 
show negligible  spectroscopic difference from Galactic counterparts. 
The situation is less 
clear for the SMC, since most WR stars are complicated by binarity. Sk41
is a possible exception, whose properties compare closely with 
Galactic or LMC counterparts (Crowther 2000).
Further progress requires detailed study of apparently single WR stars
beyond the Magellanic Clouds (e.g. Smartt et al. 2000), especially in 
nearby metal poor galaxies such as IC10 (Massey \& Armandroff 1995).

\section{Mass-loss diagnostics in yellow and red supergiants}

Various techniques have been applied to quantify mass-loss in
yellow (FG) and red (KM) supergiants, whose wind driving mechanism remains
uncertain. UV/optical observations of 
binaries with an early-type dwarf companion permit analysis of the 
red-supergiant chromosphere and wind in absorption
(Bennett et al. 1996). Radio continua
measurements have provided upper limits to mass-loss
rates of F-supergiants (Drake \& Linsky 1986) and M-supergiants (e.g. Harper
et al. 2001). Alternatively, sub-mm CO emission or mid-IR 
dust emission permit mass-loss determinations in cool supergiants.
CO emission is a reliable probe for AGB stars, although the UV
radiation field from supergiant chromospheres could lead to a 
different CO emission per unit mass loss from star to star.
Radial pulsations are generally assumed to provide the 
mechanism of initiating mass-loss in M supergiants, as evidenced from 
circumstellar dust shells (Bowers et al. 1983), with some notable
exceptions -- VY CMa (M5\,Ia) and $\rho$ Cas (F8\,Ia$^{+}$).
Outflow velocities of RSG are typically  10--40 km\,s$^{-1}$ 
(Jura \& Kleinmann 1990).

\begin{figure}[ht]
\epsfxsize=10.cm \epsfbox[020 280 450 540]{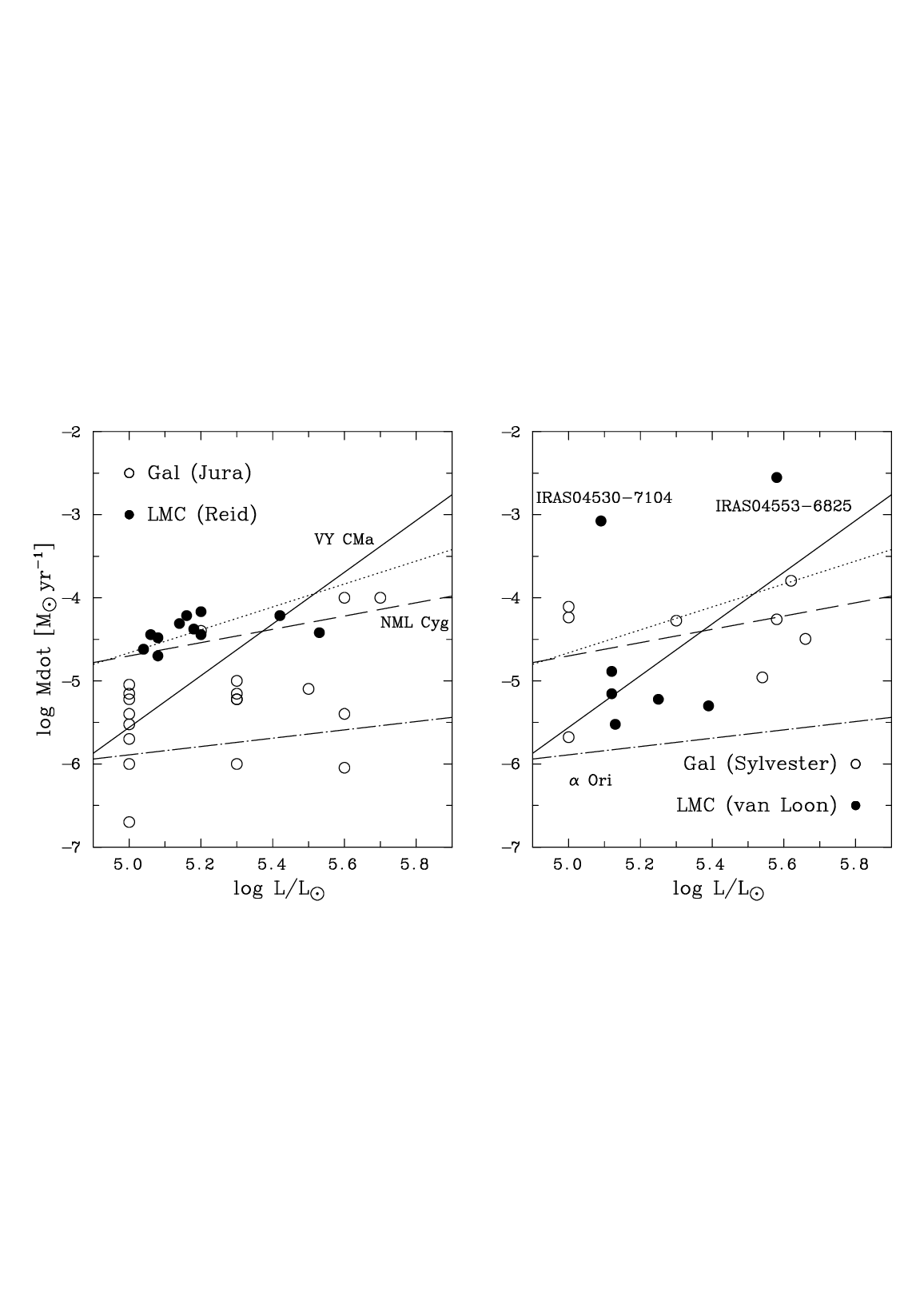}
\dblcaption{Mass-loss rates derived for Galactic (open circles) RSG
by Jura \& Kleinmann (1999) and LMC (filled circles) RSG by Reid et al.
(1990) using the Jura (1986) formula, including predictions from de Jager
et al. (1988, solid) and Reimers (1975, dot-dash) 
for $T_{\rm eff}$=3,500K. Fits to Reid et al. data are 
by Vanbeveren et al. (1998, dashed) and Salashnich et al. (1999, dotted).}
{Mass-loss rates derived from modelling
mid-IR spectroscopy of Galactic (open circles) and LMC (filled
circles) RSG by Sylvester et al. (1998) and 
van Loon et al. (1999). Distances to Galactic RSG follow
Sylvester et al. except for W Per and $\mu$ Cep, which have
been adjusted to distances derived by Garmany \& Stencel (1992). 
Lines are as described in the adjacent caption.}
\label{rsg}
\end{figure}

Since we are concerned with massive stars in this review, 
we shall restrict our discussion to the most luminous 
`hypergiants' (de Jager 1998), with log ($L/L_{\odot})>$5. 
Studies have largely focussed on M supergiants, with only
highly unusual FG supergiants studied in detail
(e.g. $\rho$ Cas, HR\,8752, IRC+10420). The latter is widely
interpreted as a RSG rapidly moving blueward (Jones et al. 1993), which
has recently experienced huge variability in mass-loss rate,
10$^{-2}$ to 10$^{-4} M_{\odot}$ yr$^{-1}$,
via analysis of its circumstellar dust shell (Bl\"{o}cker et al. 1999).

Jura \& Kleinmann (1990)
applied the `approximate' Jura (1986) formula for intermediate mass
stars to IRAS 60$\mu$m photometry of nearby yellow and red supergiants to
estimate mass-loss rates. Reid et al. (1990) followed the same approach for 
a sample of LMC RSG, which have been used by Vanbeveren et al. (1998) and 
Salasnich et al. (1999) to provide `empirical' calibrations for evolutionary
calculations, as illustrated in Fig.~2. Jura \& Kleinmann (1990) data 
for Galactic RSG reveal systematically lower rates, except for VY CMa and 
NML Cyg. Predictions from the empirical calibration of 
de Jager et al. (1988, solid) and the formulation of Reimers (1975, dot-dash)
are also illustrated, for an assumed $T_{\rm eff}$=3,500K ($\sim$M3\,I). 
Although the scatter is large, most Galactic data lie between the 
Reimers (1975) and de Jager et al. (1988) calibrations.

Recently, van Loon et al. (1999) have modelled ISO spectroscopy and 
photometry of a similar LMC sample to Reid et al., and 
found rates which are 5--10 times {\it lower}. 
Meanwhile, Sylvester et al. (1998) have used the 
10$\mu$m silicate emission feature for a sample of Galactic M supergiants, 
following the  Skinner \& Whitmore (1988) relation, which was calibrated 
against sub-mm CO determinations. Their determinations revealed
{\it higher} mass-loss rates than Jura \& Kleinmann (1990). 
From this combined dataset, presented in Fig.~3, there is little evidence
of a luminosity dependence of mass-loss. The Vanbeveren et al.
(1998) and Salasnich et al. (1999) calibrations match Galactic data
reasonably well, except for $\alpha$ Ori (M2.2\,Iab), which probably
has the most secure determination. In contrast with Fig.~2, the
majority of the LMC data indicate lower mass-loss rates than Galactic RSG, 
hinting  at a metallicity dependence. However,  van Loon et al. 
identified two LMC RSG with exceptionally high mass-loss rates of order 
10$^{-3} M_{\odot}$yr$^{-1}$  (IRAS 04553-6825 and IRAS 04530-7104)
which are unmatched in our Galaxy. 

Overall, the level of consistency between different 
IR/sub-mm techniques remains embarrassingly poor. Sylvester et al. 
(1998), using the power in 10$\mu$m silicate emission (calibrated against
earlier CO line data), derived a  mass-loss rate for 
$\mu$ Cep (M2\,Ia) which was 25 times {\it higher} than 
Jura \& Kleinmann (1990) obtained from 60$\mu$m dust emission, 
or 250 times {\it higher} than Josselin et al. (2000) based on an 
alternative CO line calibration of Olofsson et al. (1993). Indeed, 
VY CMa has values ranging over a factor of twenty -- 
3.5$\times 10^{-4}  M_{\odot}$yr$^{-1}$ (Stanek et al. 1995) to
1.6$\times 10^{-5} M_{\odot}$ yr$^{-1}$ (Josselin et al. 2000). 

In summary, the mass-loss rates for RSG are
by far the most controversial when recommending empirical calibrations
for evolutionary calculations. The majority of mass-loss estimates lie
between the Reimers (1975) and Vanbeveren et al. (1998) calibrations.
This is illustrated in Table~\ref{summary}, where we compare rates 
assumed in the recent  evolutionary model of Meynet \& Maeder (2000) 
for an initially rapidly rotating 40$M_{\odot}$ star with 
`empirical' calibrations of Kudritzki \& Puls (2000) for OBA stars, 
Vanbeveren et al. (1998) for RSG and Nugis \& Lamers (2000) for 
WR stars. Agreement is very good, except for the brief, low mass-loss 
B supergiant phase and {\it highly uncertain} high mass-loss RSG phase. 
Continuing studies into the latter are highly desirable. Typical 
outflow velocities for each evolutionary phase are also indicated in 
Table~\ref{summary}.

\begin{table}[ht]
\caption{Summary of wind properties during various phases during the
evolution of a rotating $\sim$40$M_{\odot}$ star 
($V_{\rm init}=$300 km\,s$^{-1}$) at solar metallicity 
(following Meynet \& Maeder 2000). Two sets of mass-loss rates
are given - those adopted by Meynet \& Maeder, $\dot{M}^{\rm MM}$, and
those taken from recent empirical calibrations, $\dot{M}^{\rm emp}$.} 
\label{summary}
\begin{center}
\begin{tabular}{crrcllr} 
\hline
\it Age & \it Mass & \it $\log L$ &  \it Sp Type 
& \it $\dot{M}^{\it MM}$ & \it $\dot{M}^{\it emp}$ &\it 
$v_{\infty}$ \cr
\it Myr & \it $M_{\odot}$ & \it $L_{\odot}$ & &\multicolumn{2}{c}{\it 
10$^{-6} M_{\odot}$yr $^{-1}$} & \it km\,s$^{-1}$ \cr
 \hline
0.0&40&5.4&O4\,V&\phantom{11}1.2$^a$&\phantom{11}0.4$^e$&3000$^h$\cr
3.7&35&5.5&O7\,I&\phantom{11}2.1$^a$&\phantom{11}2.1$^e$&2100$^h$\cr
5.1&31&5.7&B3\,I&\phantom{11}7.6$^a$&\phantom{11}0.7$^e$& 490$^i$\cr
5.1&31&5.8&A0\,I&\phantom{1}22.$^b$&\phantom{1}14.$^e$& 170$^i$\cr
5.1&29&5.8&M0\,I&          400.$^b$&\phantom{1}86.$^f$& 30$^j$\cr
5.3&18&5.7&WNL  &\phantom{1}30.$^c$&\phantom{1}26.$^g$& 750$^h$\cr%
5.3&16&5.6&WNE  &\phantom{1}25.$^d$&\phantom{1}31.$^g$&2000$^h$\cr
5.4&13&5.5&WCE  &\phantom{1}15.$^d$&\phantom{1}28.$^g$&2500$^h$\cr
  \hline 
  \end{tabular}
  \end{center}
\begin{tablenotes}
($a$) Lamers \& Cassinelli (1996); ($b$) de Jager et al. (1988); ($c$) Nugis et al. (1988);
($d$) Langer (1989) corrected for clumping (Schmutz 1997); ($e$) Kudritzki \&
Puls (2000); ($f$) Vanbeveren et al. (1998); ($g$) Nugis \& Lamers (2000); ($h$)
Prinja et al. (1990); ($i$) Lamers et al. (1995); ($j$) Jura \& Kleinmann (1990)
\end{tablenotes} 
 \end{table}

\begin{acknowledgments}
Many thanks to the hospitality of JILA, University of Colorado at Boulder where
this review was written, especially Prof P.S. Conti. Financial support
is provided by a Royal Society URF. Thanks to
Prof M.J. Barlow and Dr  G. Harper for useful discussions and to Dr G. Meynet
for providing recent grids of evolutionary calculations.
\end{acknowledgments}


\begin{chapthebibliography}{1}
\bibitem{} Achmad, L., Lamers, H.J.G.L.M. \&  Pasquini, L., 1997, 
A\&A {\bf 320}, 196 
\bibitem{} Barba,~R.H.,~Niemela,~V.S.,~Baume,~G. \& Vazquez,~R.A., 1995, 
ApJ {\bf 446}, L23 
\bibitem{} Barlow, M.J. \& Cohen, M., 1977,
ApJ {\bf 213}, 737 
\bibitem{} Bennett, P.D., Harper, G.M., Brown, A. \& Hummel, C.A., 1996, 
ApJ {\bf 471}, 454 
\bibitem{} Bieging, J.H., Abbott, D.C. \& Churchwell, E.B., 1989, 
ApJ {\bf 340}, 518 
\bibitem{} Bl\"{o}cker, T., Balega, Y., Hofmann, K.-H. et al. 1999, 
A\&A {\bf 348}, 805 
\bibitem{} Bohannan, B. \& Crowther, P.A., 1999, 
ApJ {\bf 511}, 374 
\bibitem{} Bowers, P.F., Johnston, K.J. \& Spencer J.H., 1983, 
ApJ {\bf 274}, 733 
\bibitem{} Bjorkman, J.E. \& Cassinelli, J.P., 1993, 
ApJ {\bf 409}, 429 
\bibitem{} Castor J.I., Abbott D.C. \& Klein, R.I., 1975, 
ApJ {\bf 195}, 157 
\bibitem{} Chapman J.M., Leitherer, C., Koribalski, B., Bouter, R. \& Storey, M, 1999, 
ApJ {\bf 518}, 890 
\bibitem{} Chlebowski, T., Harnden, F.R.Jr. \& Sciortino, S., 1989, 
ApJ {\bf 341}, 427 
\bibitem{} Crowther, P.A., 1997, in Proc. LBVs: Massive Stars in Transition,
ASP Conf. Ser. {\bf 120} (eds. Nota, A., Lamers, H.J.G.L.M.) p.51 
\bibitem{} Crowther, P.A., 2000, 
A\&A {\bf 356}, 191 
\bibitem{} Crowther, P.A. \& Smith, L.J., 1997, 
A\&A {\bf 320}, 500 
\bibitem{} Davidson, K. \& Humphreys, R. M., 1997, 
ARA\&A {\bf 35}, 1 
\bibitem{} de Jager, C., 1998, 
A\&A Rev {\bf 8}, 145 
\bibitem{} de Jager, C., Nieuwenhuijzen, H. \& van der Hucht, K.A., 1988, 
A\&AS {\bf 72}, 259 
\bibitem{} Dessart, L., Crowther, P.A., Hillier, D.J. et al. 2000, 
MNRAS {\bf 315}, 407 
\bibitem{} Drake, S.A. \& Linsky, J.L., 1986,
AJ {\bf 91}, 602 
\bibitem{} Drake, S.A. \& Linsky, J.L., 1989, 
AJ {\bf 98}, 1831 
\bibitem{} Drissen,  L., Roy, J.-R. \& Robert, C., 1997, 
ApJ {\bf 474}, L35 
\bibitem{} Drissen, L., Crowther, P.A., Smith, L.J. et al. 2001, 
ApJ in press (astro-ph/0008221)
\bibitem{} Eenens, P.R.J. \& Williams, P.M., 1994, 
MNRAS {\bf 269}, 1082 
\bibitem{} Feldmeier, A., 1995, 
A\&A {\bf 299}, 523 
\bibitem{} Friend, D.B. \& Abbott, D.C., 1986, 
ApJ {\bf 311}, 701 
\bibitem{} Fullerton, A.W., Gies, D.R. \& Bolton, C.T., 1996, 
ApJS {\bf 103}, 475 
\bibitem{} Fullerton, A.W. \& Najarro, F., 1998, in Proc.  Boulder-Munich II: 
Properties of hot, luminous stars, ASP Conf Ser. {\bf 131} (I.D. Howarth, ed.), p.47
\bibitem{} Fullerton~A.W.,~Crowther,~P.A.,~De~Marco,~O.~et al., 2000,
ApJ~{\bf 538},~L43 
\bibitem{} Garmany, C.D. \& Conti, P.S., 1985, 
ApJ {\bf 293}, 407 
\bibitem{} Garmany, C.D. \& Stencel, R.E., 1992, 
A\&AS {\bf 94}, 211 
\bibitem{} Groenewegen, M.A.T., Lamers, H.J.G.L.M. \& 
Pauldrach, A.W.A., 1989,
A\&A {\bf 221}, 78 
\bibitem{} Hamann, W-.R. \& Koesterke, L., 2000, 
 A\&A {\bf 360}, 647 
\bibitem{} Harper, G.M.,  Brown, A. \& Lim, J., 2001, 
ApJ submitted
\bibitem{} Harries, T.J., Hillier, D.J. \& Howarth, I.D., 1998, 
MNRAS {\bf 296}, 1072 
\bibitem{} Haser,~S.M.,~Pauldrach,~A.W.A.,~Lennon,~D.J.~et~al., 1998,
A\&A~{\bf 330},~285
\bibitem{} Hillier, D.J., 1991, 
A\&A {\bf 247}, 455 
\bibitem{} Hillier, D.J., Crowther, P.A., Najarro, F. \& 
Fullerton, A.W., 1998,
 A\&A {\bf 340}, 483 
\bibitem{} Hillier, D.J., Davidson K., Ishibashi, K. \& Gull T., 2000, 
ApJ in press
\bibitem{} Hillier D.J. \& Miller D., 1998, 
ApJ {\bf 496}, 407 
\bibitem{} Hillier D.J. \& Miller D., 1999, 
ApJ {\bf 519}, 354 
\bibitem{} Howarth, I.D., Siebert, K.W., Hussain, G.A.J. \& 
Prinja, R.K., 1997,
MNRAS {\bf 284}, 265 
\bibitem{} Jones, T.J., Humphreys, R.M., Gehrz, R.D. et al. 1993, 
ApJ {\bf 411}, 323 
\bibitem{} Josselin, E., Blommaert J.A.D.L., Groenewegen, M.A.T., 
Omont, A. \& Li, F.L., 2000, 
A\&A {\bf 357}, 225 
\bibitem{} Jura, M. 1986,
ApJ {\bf 303}, 327 
\bibitem{} Jura, M. \& Kleinmann, S.G., 1990, 
ApJS {\bf 73}, 769 
\bibitem{} Kudritzki, R-.P., Puls, J., Lennon, D.J. et al. 1999, 
A\&A {\bf 350}, 970 
\bibitem{} Kudritzki, R-.P. \& Puls, J. 2000,
ARA\&A {\bf 38}, 613
\bibitem{} Lamers, H.J.G.L.M., Cerruti-Sola, M. \& Perinotto, M., 1987, 
ApJ {\bf 314}, 726 
\bibitem{} Lamers, H.J.G.L.M. \& Leitherer, C., 1993,  
ApJ {\bf 412}, 771 
\bibitem{} Lamers, H.J.G.L.M., Snow, T.P. \& Lindholm, D.M., 1995, 
ApJ {\bf 455}, 269 
\bibitem{} Lamers, H.G.J.L.M. \& Cassinelli, J.P., 1996, in 
Proc. From Stars to Galaxies: The Impact of Stellar Phjysics on Galaxy Evolution, ASP Conf. Ser. {\bf 98} (eds. Leitherer C., Fritze-vob-Alvensleben, U., Huchra, J.), p.162
\bibitem{} Lamers, H.G.J.L.M. \& Cassinelli, J.P., 1999, Introduction to Stellar Winds, CUP, Cambridge
\bibitem{} Lamers, H.G.J.L.M., Haser, S., de Koter, A. \&
Leitherer, C., 1999,
ApJ {\bf 516}, 872 
\bibitem{} Langer, N. 1989, 
A\&A {\bf 220}, 135 
\bibitem{} Leitherer, C., 1997, in Proc. LBVs: Massive Stars in Transition,
ASP Conf. Ser. {\bf 120} (eds. Nota, A., Lamers, H.J.G.L.M.) p.58
\bibitem{} Leitherer, C., Chapman, J.M. \& Koribalski, B., 1997 
ApJ {\bf 481}, 898 
\bibitem{} Lepine, S., Dalton, M.J., Moffat, A.F.J. et al. 2000, 
AJ in press
\bibitem{} Lucy, L.B. \& White, R.L. 1980, 
ApJ {\bf 241}, 300 
\bibitem{} Massey, P. \& Armandroff, T.E., 1995,
AJ {\bf 109}, 2470 
\bibitem{} Meynet, G., Maeder, A., Schaller, G., Schaerer, D. \& 
Charbonnel, C., 1994, 
A\&AS {\bf 103}, 97 
\bibitem{} Meynet, G., \& Maeder, A., 2000, 
A\&A {\bf 361}, 101 
\bibitem{} Morris P.W., van der Hucht, K.A., Crowther, P.A. et al. 2000, 
A\&A {\bf 353}, 624 
\bibitem{} Najarro, F., Hillier, D.J., Kudritzki, R-.P. et al. 1994,
A\&A {\bf 285}, 573 
\bibitem{} Najarro, F., Hillier, D.J. \& Stahl, O. 1997, 
A\&A {\bf 326}, 1117 
\bibitem{} Nugis, T., Crowther, P.A. \& Willis, A.J. 1998,
A\&A {\bf 333}, 956 
\bibitem{} Nugis, T. \& Lamers, H.J.G.L.M. 2000,
A\&A {\bf 360}, 227 
\bibitem{} Panagia, N. \& Felli, M. 1975, 
A\&A {\bf 39}, 1 
\bibitem{} Pauldrach A.W., Puls J. \& Kudritzki R-.P. 1986,
A\&A {\bf 164}, 86 
\bibitem{} Peimbert, M., Peimbert, A. \& Ruiz, M.T. 2000, 
ApJ {\bf 541}, 688 
\bibitem{} Petrenz, P. \& Puls, J. 2000, 
A\&A {\bf 358}, 956 
\bibitem{} Prinja,~R.K., 1992,~in~Proc.~Nonisotropic~and~Variable~Outflows~from~Stars, ASP Conf. Ser. {\bf 22}, (eds. Drissen, L., Leitherer, C., Nota, A.) p.167
\bibitem{} Prinja, R.K., Barlow, M.J. \& Howarth, I.D., 1990, 
ApJ {\bf 361}, 607 
\bibitem{} Prinja, R.K. \& Crowther, P.A. 1998,
MNRAS {\bf 300}, 828 
\bibitem{} Prinja, R.K., Stahl, O., Kaufer A. et al. 2000,
A\&A submitted
\bibitem{} Puls, J., Kudritzki, R-.P., Herrero, A. et al. 1996,
A\&A {\bf 305}, 171 
\bibitem{} Reid, N., Tinney, C. \& Mould, J., 1990, ApJ {\bf 348}, 98-119
\bibitem{} Reimers, D., 1975, In: Problems in stellar atmospheres and envelopes,
Springer-Verlag (New York), p.229 
\bibitem{} Olofsson, H., Eriksson, K., Gustafsson, B. \& 
Carlstrom, U. 1993, 
ApJS {\bf 87}, 267 
\bibitem{} Owocki, S.P., Castor, J.I. \& Rybicki, G.B. 1988,
ApJ {\bf 335}, 914 
\bibitem{} Owocki, S.P., Cranmer, S.R. \& Gayley, K.G. 1996,
ApJ {\bf 472}, L115 
\bibitem{} Salasnich, B., Bressan, A. \& Choisi, C. 1999,
A\&A {\bf 342}, 131 
\bibitem{} Schaerer, D. \& Maeder, A. 1992, 
A\&A {\bf 263}, 129 
\bibitem{} Schmutz, W. 1997, 
A\&A {\bf 321}, 268 
\bibitem{} Skinner, C.J. \& Whitmore, B. 1988,
MNRAS {\bf 231}, 169 
\bibitem{} Smartt, S.J., Crowther, P.A., Dufton, P.L. et al. 2000
MNRAS submitted (astro-ph/0009156)
\bibitem{} Smith, L.F. \& Maeder, A., 1991,
A\&A {\bf 241}, 77 
\bibitem{} Sylvester, R., Skinner, C.J. \& Barlow, M.J. 1998,
MNRAS {\bf 301}, 1083 
\bibitem{} Stanek, K.Z., Knapp, G.R., Young, K. \& Phillips, R.G. 1995,
ApJS {\bf 100}, 169 
\bibitem{} van Loon, J.Th., Groenewegen, M.A.T., de Koter, A. et al. 1999, 
A\&A {\bf 351}, 559 
\bibitem{} Vanbeveren D., De Loore, C. \& Van Rensbergen, W. 1998,
A\&A Rev. {\bf 9}, 63 
\bibitem{} Vink, J.S., de Koter, A. \& Lamers, H.J.G.L.M. 1999, 
A\&A {\bf 350}, 181 
\bibitem{} Vink, J.S., de Koter, A. \& Lamers, H.J.G.L.M. 2000, 
A\&A in press
\bibitem{} von Zeipel, H., 1924, MNRAS {\bf 84}, 665
\bibitem{} Walborn, N.R., Nichols-Bohlin, J. and Panek, R.J., 1985, 
IUE Atlas of O-type spectra from 1200 to 1900\AA. NASA RP 1155.
\bibitem{} Walborn, N.R., Lennon, D.J., Haser, S., Kudritzki, R-.P. \& 
Voels, S.A., 1995,
PASP {\bf 107}, 104 
\bibitem{} Waldron, W.L. \& Cassinelli, J.P. 2000, ApJ Letters submitted
\bibitem{} Wickramasinghe,~N.C.,~Donn,~B.D.~\&~Stecher~T.P.,~1966,~ApJ~{\bf 146},~590 
\bibitem{} Wright, A.E. \& Barlow, M.J. 1975,
MNRAS {\bf 170}, 41 
\end{chapthebibliography}
\end{document}